\documentclass[10pt,twoside]{article}
\usepackage{amssymb,amsmath}
\def\volnum{Manuscrit}

\usepackage{aflbcours}
\pdebut{1}
\def\pacs#1{\LP P.A.C.S.: #1}
\title{Beables for Quantum Electrodynamics}
\titleshort{Beables for Quantum Electrodynamics}
\author{Samuel Colin}
\authorshort{S. Colin}
\address{University of Louvain-la-Neuve, FYMA\\
2 chemin du cyclotron, B-1348 Louvain-la-Neuve, Belgium}

\begin{document}
\maketitle

\vskip 1cm
\begin{abstract}
ABSTRACT. We show that it is possible to obtain a realistic and deterministic model, based on a previous work
of John Bell, which reproduces the experimental predictions of the orthodox interpretation of quantum electrodynamics.
\end{abstract}
\section{Introduction}
First, we would like to recall why the measurement postulate has been introduced in the orthodox interpretation of the 
quantum theory. Let us assume that a system of particles is completely described by a wave-function, that always evolves according 
to a linear equation (such as the Schr\"odinger equation). We plan to measure an observable $\mathcal{A}$ of the system. To measure the observable 
$\mathcal{A}$, one has to build an apparatus that correlate the eigenstates of $\mathcal{A}$ to position eigenstates of the apparatus, so that 
any measurement is finally a measurement of positions. For example, when the spin component of a spin-$\frac{1}{2}$ particle is measured, 
as in the Stern-Gerlach experiment, we finally measure a point on a screen, so a 
position (either in the upper half-plate, either in the lower half-plate). This remark is important since it gives a 
privileged role to position, and so superpositions get an absolute meaning. Now if we prepare the system in a superposition 
of eigenstates of $\mathcal{A}$, and if we send it towards the measuring apparatus, we can obtain superpositions of macroscopic objects 
(measuring devices, cats, humans, and so on), on account of the 
linear character of the equation of motion. But we do 
not observe those macroscopic superpositions, but eigenstates of the observable being measured, with probabilities 
given by Born's law. In order to account for the experimental results, 
the measurement postulate is introduced:
\begin{quote}
When an observer measure the observable $\mathcal{A}$ of the system, he founds one of the eigenvalues of $\mathcal{A}$, 
with probability given by Born's law. Because of the repeatability of experiments, the wave-function of the system collapses 
to the corresponding eigenstate when the measurement occurs. 
\end{quote}
The measurement postulate introduces a boundary between two worlds: a quantum world, made of systems, described by wave-functions, 
and a classical world, made of observers, described by positions. What is wrong with that? According to the orthodox 
interpretation, there is no element of reality in the quantum world: the particles of the quantum world do not 
have any probability of being anywhere (they do not exist). Elements of the classical world 
do have such probabilities: they exist (their positions are elements of reality). And the orthodox 
interpretation offers no explanation for that. People whose main concern is quantum gravity must face this kind 
of problem, for it is meaningless to speak of a wave-function of the universe in the orthodox interpretation. 

To get rid of the frontier between the classical and the quantum worlds, one could say that the world is described by 
a wave-function and by the positions of the particles that it contains. There is a theory, built along 
these lines; it is the non-relativistic de Broglie-Bohm pilot-wave theory. Pilot-wave theory is a realistic theory, in so far as the
positions of the particles exist and are simply revealed by position measurements (the positions are beables, a term coined by
Bell). If there are $n$ particles, the universe is thus completely described by the couple $(\vec{X}(t),\Psi(t,\vec{X}))$, where
$\vec{X}$ is a point in a configuration space of dimension $3n$. $\Psi(t,\vec{X})$ evolves according to the Schr\"odinger equation,
whereas the equation of motion for $\vec{X}(t)$ is such that if we consider a set of universes with the same wave-function and
initial configurations chosen according to the probability density $|\Psi(t_0,\vec{X})|^2$, then the final configurations will be
distributed according to $|\Psi(t,\vec{X})|^2$ (for any later time $t$). So the non-relativistic pilot-wave theory gives the same 
experimental predictions as the orthodox interpretation of the non-relativistic quantum theory, without relying on concepts such as 
systems and observers.

Pilot-wave theory is weakly non-local; that means that the equation of motion for the positions of the particles is non-local, but that does 
not lead to any supra-luminal signaling. Now, the widespread claim is that those interpretations
are ruled out by Bell's inequality (\cite{bell2}, chap 7) and the experiments that have been carried out later. If fact local ones are ruled out, but
since non-locality is commonly claimed to be unacceptable (even weak non-locality, which does not lead to paradoxical situations),
it is said that any hidden-variables theory is incompatible with the quantum theory. That is the wrong way to present the
theoretical situation. To present it correctly, we have to return to the EPR paradox \cite{epr}, which in essence says that some quantum
correlations cannot be explained in a local way, unless we say that the quantum theory is incomplete. Since local
hidden-variables theories are ruled out by Bell's inequality and experiments, the only way to explain quantum correlations is to
revert to non-locality (which is in fact hidden in the collapse postulate). Then, to suppress the ill-defined measurement
postulate, it is preferable to interpret the quantum theory by a non-local hidden-variables theory, such as the non-relativistic pilot-wave theory. 
It is worth underlining that Bell, who is often credited of the refutation of hidden-variables theories, has been one of the 
main advocates of the pilot-wave theory (\cite{bell2}, chap. 11, 14, 15 and 17).

The question that comes to mind is to know if the quantum field theory can also be interpreted as a non-local hidden-variables
theory. At the time Bell wrote his paper \cite{bell1} (or \cite{bell2}, chap. 19), Bohm had already shown that it was possible to build a realistic
interpretation of any bosonic quantum field theory \cite{bohm1}. To achieve that goal, Bohm took the field as the beable, however
he was not able to do the same for fermions. The aim of Bell was then to show that it was also possible to build a realistic
interpretation of any fermionic quantum field theory, along the pilot-wave ideas \cite{bell1}. Bell managed doing so but he took a different
beable: the fermion number density. It is slightly different from the non-relativistic pilot-wave theory, whose beables are the
positions of the particles. The model is also formulated on a spatial lattice (space is discrete but time remains continuous).
His model is stochastic, but he suspected that the theory would become deterministic in the continuum limit. We have shown 
\cite{colin} that it is indeed the case, so that there is a deterministic Bell model.

In this article, we consider in more details the deterministic Bell model for quantum electrodynamics. 
First, we will give a brief presentation of the non-relativistic pilot-wave theory, 
since the deterministic Bell model is in fact very similar to it. Then, we will give a presentation of the Dirac quantum field theory, axed around the
notion of negative-energy electrons. In that approach, there are only particles of charge $-e$ but of positive and negative energy:
their total number (fermion-number) is conserved. We will see that this approach is a convenient one to study localized properties
(such as fermion-number density). The reinterpretation of the theory is then made in terms of positrons, as holes in the Dirac sea. In section 
\ref{sec:dbm}, we build the deterministic Bell model for the free Dirac quantum field theory. 
In the next section, the results are generalized to quantum electrodynamics. 

\section{The non-relativistic pilot-wave theory}
Let us consider a system containing $n$ non-relativistic electrons. In the orthodox interpretation, these electrons are 
completely described by a wave-function $\Psi_{s_1\ldots s_n}(t,\vec{x}_1,\ldots,\vec{x}_n)$, which is the solution of the Schr\"odinger 
equation ($s_1,\ldots,s_n$ are spin-component indices). 
The probability density to observe the system in a configuration $(\vec{x}_1,\ldots,\vec{x}_n)$ at time $t$ is given by 
\begin{equation*}
\rho(t,\vec{x}_1,\cdot\cdot,\vec{x}_n)=
\sum_{s_1=1}^{s_1=2}\cdot\cdot\sum_{s_n=1}^{s_n=2}\Psi^*_{s_1\ldots s_n}(t,\vec{x}_1,\ldots,\vec{x}_n)\Psi_{s_1\ldots s_n}(t,\vec{x}_1,\ldots,\vec{x}_n)~.
\end{equation*}
From the relation
\begin{equation*}
\int d^3\vec{x}_1\ldots d^3\vec{x}_n \rho(t,\vec{x}_1,\ldots,\vec{x}_n)=1
\end{equation*}
and with the help of the Schr\"odinger equation, it can be shown, at least in the usual cases, 
that there exist currents $\vec{j}_1, \ldots,\vec{j}_n$ such that 
\begin{equation}\label{ce}
\frac{\partial\rho(t,\vec{X})}{\partial t}+\sum_{k=1}^{k=n}\vec{\nabla}_{\vec{x}_k}\cdot\vec{j}_k(t,\vec{X})=0~.
\end{equation}
We use $\vec{X}$ as a short notation for $(\vec{x}_1,\ldots,\vec{x}_n)$. 

The main idea of the pilot-wave theory is to say that the positions of the electrons are in fact elements 
of reality (or beables); that means that the electrons have definite positions $\vec{x}_1(t),\ldots,\vec{x}_n(t)$
that are simply revealed by position measurements. So the complete description 
of a system at time $t$ would involve the wave-function $\Psi_{s_1\ldots s_n}(t,\vec{x}_1,\ldots,\vec{x}_n)$, but also the $n$ position 
variables $\vec{x}_1(t),\ldots,\vec{x}_n(t)$. To complete the theory, one must give the equations of motion for these objects. 
For the wave-function, 
the Schr\"odinger equation is retained. As far as the position variables are concerned, the only constraint on the equation of motion 
is that the predictions of the orthodox interpretation of the quantum theory are reproduced. Is there a general form for 
that equation of motion? There are only two basic objects defined at time $t$: the wave-function of the system and the 
position of the system ($\vec{X}(t)=(\vec{x}_1(t),\ldots,\vec{x}_n(t))$). 
So the velocity of the system would be fixed by those objects. In other words, the equation of motion is a velocity-law:
\begin{equation*}
\vec{X}(t+dt)=\vec{X}(t)+\vec{V}^{\Psi}(t,\vec{X})|_{\vec{X}=\vec{X}(t)}~,
\end{equation*}
where $\vec{V}^{\Psi}(t,\vec{X})=(\vec{v}^{\Psi}_1(t,\vec{X}),\ldots,\vec{v}^{\Psi}_n(t,\vec{X}))$.

Let us see what is the constraint imposed on the velocity-law by requiring that the pilot-wave theory is in agreement with the orthodox quantum 
theory. Let us define $r(t,\vec{X})$ as the probability density for the system to be in configuration 
$\vec{X}$ at time $t$. The probability density is defined in the usual way: we consider a set of systems, labeled by an index $\alpha$, with 
the same wave-function but different realizations of the beable $\vec{X}_\alpha(t)$. Then the probability density is defined over 
these many experiments $\alpha$. If the positions are simply revealed by position measurements, the probability density for the system 
to be observed in a configuration $\vec{X}$ is equal to the probability density for the system to be in configuration $\vec{X}$. Hence, 
to reproduce the predictions of the orthodox quantum theory, the following relation 
\begin{equation*}
\rho(t,\vec{X})=r(t,\vec{X})
\end{equation*}
must be satisfied.
Let us assume that this is true for some initial time $t_0$; then the relation that must be satisfied is 
\begin{equation}\label{constraint}
\frac{\partial\rho(t,\vec{X})}{\partial t}=\frac{\partial r(t,\vec{X})}{\partial t}~.
\end{equation}
We first consider the right-hand part. A consequence of the conservation of the total probability 
\begin{equation*}
\int d^{3n}\vec{X}r(t,\vec{X})
\end{equation*}
is the existence of a continuity equation 
\begin{equation}\label{pwce}
\frac{\partial r(t,\vec{X})}{\partial t}+\vec{\nabla}_{\vec{X}}\cdot(r(t,\vec{X})\vec{V}^{\Psi}(t,\vec{X}))=0~.
\end{equation}
With the help of eq. (\ref{ce}) and eq. (\ref{pwce}), eq. (\ref{constraint}) can be rewritten as 
\begin{align*}
\vec{\nabla}_{\vec{X}}\cdot(r(t,\vec{X})\vec{V}^{\Psi}(t,\vec{X}))=&\sum_{k=1}^{k=n}\vec{\nabla}_{\vec{x}_k}
\cdot(r(t,\vec{X})\vec{v}^{\Psi}_k(t,\vec{X}))&\\
=&\sum_{k=1}^{k=n}\vec{\nabla}_{\vec{x}_k}\cdot\vec{j}_k(t,\vec{X})~.&
\end{align*}
Taking 
\begin{equation}\label{vlaw}
\vec{v}^{\Psi}_k(t,\vec{X})=\frac{\vec{j}_k(t,\vec{X})}{r(t,\vec{X})}=\frac{\vec{j}_k(t,\vec{X})}{\rho(t,\vec{X})}~,
\end{equation}
the predictions of the orthodox quantum theory are regained. A complete study of the pilot-wave theory can be found in \cite{holland}.
\section{\label{sec:dqft}The free Dirac quantum field theory}
Our aim now is to show that the idea of positive and negative energy electrons is convenient for the study of localized properties, such as fermion-number 
density (or charge density). 
\subsection{The Dirac equation in first quantization}
The classical Dirac equation is
\begin{equation*}
i\frac{\partial\psi(t,\vec{x})}{\partial
t}=-i\vec{\alpha}\cdot\vec{\nabla}\psi(t,\vec{x})+m\beta\psi(t,\vec{x})~,
\end{equation*}
where the hermitian matrices $\beta$ and $\alpha^j$ ($j=1,2,3$)
satisfy the relations
\begin{align*}
&\{\alpha^j,\beta\}=0& &\{\alpha^j,\alpha^k\}=2\delta^j_k&
&\beta^2=1&
\end{align*}
and where $\psi(t,\vec{x})$ is a four-component spinor field. A
particular representation of the matrices is given by the
Pauli-Dirac choice:
\begin{align*}
&\beta=\begin{pmatrix}1&0\\0&-1\end{pmatrix}&
&\vec{\alpha}=\begin{pmatrix}0&\vec{\sigma}\\\vec{\sigma}&0\end{pmatrix}~,&
\end{align*}
where the $\sigma^j$ are the usual Pauli matrices.

The Dirac equation can be rewritten in a covariant form by
introducing the $\gamma$ matrices, defined by the relations
\begin{align*}
&\gamma^0=\beta=\gamma_0& &\gamma^j=\alpha^j=-\gamma_j~.&
\end{align*}
Doing so leads to
\begin{equation*}
i\gamma^\mu\partial_\mu\psi(t,\vec{x})-m\psi(t,\vec{x})=0~.
\end{equation*}
It can also be shown that the $\gamma$ matrices satisfy the
relations $\{\gamma^\mu,\gamma^\nu\}=2g^{\mu\nu}$. Then, it is
easy to check that each component of the Dirac field is a solution
of the Klein-Gordon equation
\begin{equation*}
(\square+m^2)\psi_a(t,\vec{x})=0~,
\end{equation*}
with $a=1,2,3,4$. Since the free solutions of the K-G equation are
$e^{-iE_{\vec{p}}t}e^{i\vec{p}\cdot\vec{x}}$ and
$e^{iE_{\vec{p}}t}e^{-i\vec{p}\cdot\vec{x}}$, with
$E_{\vec{p}}=\sqrt{|\vec{p}|^2+m^2}$ and $\vec{p}\in\mathbb{R}^3$,
the general forms of the free solutions of the Dirac equation are
\begin{align*}
&u_s(\vec{p})e^{-iE_{\vec{p}}t}e^{i\vec{p}\cdot\vec{x}}&
&v_s(\vec{p})e^{iE_{\vec{p}}t}e^{-i\vec{p}\cdot\vec{x}}~,&
\end{align*}
with $s=1,2$ and $\vec{p}\in\mathbb{R}^3$. To give an
interpretation of the Dirac equation, one must find a conserved
current, whose time-component is positive; the current
$j^\mu=\bar{\psi}\gamma^\mu\psi$, where
$\bar{\psi}=\psi^\dagger\gamma^0$, is suitable. The quantity
$j^0(t,\vec{x})=\psi^\dagger(t,\vec{x})\psi(t,\vec{x})$ is thus
the probability density to find an electron at point $\vec{x}$ at
time $t$. Normalization still remains to be discussed. If the free
particle has momentum $\vec{p}_{\Sigma}$ in an inertial frame $\Sigma$,
where the universe appears to have a volume $V_{\Sigma}$, then the
following relation must be satisfied:
\begin{equation*}
\int_{V_{\Sigma}}d^3\vec{x}\psi^\dagger_{\vec{p}_{\Sigma}}(t,\vec{x})\psi_{\vec{p}_{\Sigma}}(t,\vec{x})=1~,
\end{equation*}
in every inertial frame. That means that the spinors must be
normalized to
\begin{align*}
&u^\dagger_s(\vec{p})u_s(\vec{p})=\frac{E_{\vec{p}}}{V m}&
&v^\dagger_s(\vec{p})v_s(\vec{p})=\frac{E_{\vec{p}}}{V m}~,&
\end{align*}
where $V$ is the volume of the universe in the inertial frame
where the particle is at rest.

The time component of the current is positive, a task that was impossible with the Klein-Gordon
equation. But the negative energy states are still there. Once
interactions are taken into account, their presence would lead to the
instability of the hydrogen atom, for example. To avoid this,
Dirac has assumed that all the negative energy states were occupied.
Hence a positive energy electron cannot transit to a negative
energy state, due to the Pauli exclusion principle. That state of
lowest energy is called the Dirac sea. The absence of
a negative energy state of momentum $\vec{p}$ (a hole in the Dirac
sea) would be seen as a particle of positive energy $E_{\vec{p}}$,
momentum $-\vec{p}$ and charge $e$. That led eventually to the prediction of
anti-particles known as positrons.
\subsection{The Dirac theory in second quantization}
The first step, in the construction of the corresponding quantum
field theory, is to find a real classical relativistic action,
that leads to the Dirac equation, when the variational principle
is applied on it. The following action
\begin{equation*}
S=\int d^3\vec{x}dt\mathcal{L}(t,\vec{x})=\int
d^3\vec{x}dt\bar{\psi}(t,\vec{x})[i\gamma^\mu\partial_\mu-m]\psi(t,\vec{x})
\end{equation*}
is a good candidate. It is real, up to a four-divergence. The
momenta conjugate to the fields are
\begin{align*}
&\pi_a(t,\vec{x})=\frac{\partial\mathcal{L}}{\partial\dot{\psi}_a(t,\vec{x})}=i\psi^*_a(t,\vec{x})&
&\pi^*_a(t,\vec{x})=\frac{\partial\mathcal{L}}{\partial\dot{\psi}^*_a(t,\vec{x})}=0~.&
\end{align*}
The next step is quantization; classical fields become quantum
fields, satisfying the equal-time canonical anti-commutation
relations
\begin{align}\label{car}
&\{\psi_a(t,\vec{x}),\psi^\dagger_b(t,\vec{y})\}=\delta^a_{b}\delta^3(\vec{x}-\vec{y})&
&\{\psi_a(t,\vec{x}),\psi_b(t,\vec{y})\}=0~.&
\end{align}
Since the quantum field $\psi(t,\vec{x})$ is a solution of the
Dirac equation, it is a superposition of free classical solutions
with operators as coefficients:
\begin{equation*}
\psi(t,\vec{x})=\sqrt{\frac{1}{V}}\sum_{s,\vec{p}}\sqrt{\frac{m}{E_{\vec{p}}}}
[c_s(\vec{p})u_s(\vec{p})e^{-iE_{\vec{p}}t}e^{i\vec{p}\cdot\vec{x}}+
\zeta_s(-\vec{p})v_s(\vec{p})e^{iE_{\vec{p}}t}e^{-i\vec{p}\cdot\vec{x}}]
\end{equation*}
(summation is made over all the momentum
$\vec{p}=(n_1\frac{2\pi}{L},n_2\frac{2\pi}{L},n_3\frac{2\pi}{L})$,
with $V=L^3$ and $n_1,n_2,n_3\in\mathbb{R}^3$). Taking
\begin{align*}
&\{c_s(\vec{p}),c^\dagger_r(\vec{q})\}=\delta^r_s\delta_{\vec{p}\;\!\vec{q}}&
&\{\zeta_s(\vec{p}),\zeta^\dagger_r(\vec{q})\}=\delta^r_s\delta_{\vec{p}\;\!\vec{q}}~,&
\end{align*}
and all other anti-commutators vanishing, the relations (\ref{car}) are regained.

Let us now give an
interpretation of the operators $c$ and $\zeta$. The hamiltonian
is obtained through the Legendre transformation
\begin{align*}
{H_{\mathcal{D}}}=&\int
d^3\vec{x}\psi^\dagger(t,\vec{x})(-i\vec{\alpha}\cdot\vec{\nabla}+m\beta)\psi(t,\vec{x})&\\=&
\sum_{s,\vec{p}}\sqrt{|\vec{p}|^2+m^2}(c^\dagger_s(\vec{p})c_s(\vec{p})-\zeta^\dagger_s(\vec{p})\zeta_s(\vec{p}))~.&
\end{align*}
The momentum is
\begin{equation*}
\vec{P}=\int
d^3\vec{x}\psi^\dagger(t,\vec{x})(-i\vec{\nabla})\psi(t,\vec{x})=
\sum_{s,\vec{p}}\vec{p}\;\!(c^\dagger_s(\vec{p})c_s(\vec{p})+\zeta^\dagger_s(\vec{p})\zeta_s(\vec{p}))~.
\end{equation*}
We define the fermion-number as the operator
\begin{equation*}
F=\int d^3\vec{x}\psi^\dagger(t,\vec{x})\psi(t,\vec{x})=
\sum_{s,\vec{p}}(c^\dagger_s(\vec{p})c_s(\vec{p})+\zeta^\dagger_s(\vec{p})\zeta_s(\vec{p}))~,
\end{equation*}
$\psi^\dagger(t,\vec{x})\psi(t,\vec{x})$ being the fermion-number
density. Fermion-number means number of fermions, that is number
of positive-energy electrons plus number of negative-energy
electrons. The fermion-number is conserved ($[{H_{\mathcal{D}}},F]=0$); the
corresponding current is
$j^\mu=\bar{\psi}\gamma^\mu\psi=(F,\vec{J})$ with $\partial_\mu
j^\mu=0$. Since there are only electrons of charge $-e$, the fact that the fermion-number is conserved is equivalent to the
charge conservation. The current $\vec{J}$ is
\begin{equation*}
\vec{J}=\int
d^3\vec{x}\psi^\dagger(t,\vec{x})\vec{\alpha}\psi(t,\vec{x})=
\sum_{s,\vec{p}}\vec{p}\;\!(c^\dagger_s(\vec{p})c_s(\vec{p})-\zeta^\dagger_s(\vec{p})\zeta_s(\vec{p}))~.
\end{equation*}
The charge density is $Q(\vec{x})=-eF(\vec{x})$ and the charge current is $-e\vec{J}$.

We define a vacuum
$|0_\mathcal{D}\rangle$ as a state destroyed by the operators $c$
and $\zeta$:
\begin{align*}
&c_s(\vec{p})|0_\mathcal{D}\rangle=0& &\zeta_s(\vec{p})|0_\mathcal{D}\rangle=0&
&\forall~s,\vec{p}~.&
\end{align*}
This vacuum $|0_\mathcal{D}\rangle$ is not the Dirac sea (the Dirac sea is
obtained from $|0_\mathcal{D}\rangle$ by filling all the negative-energy
states). Then $c^\dagger_s(\vec{p})|0_\mathcal{D}\rangle$ and $\zeta^\dagger_s(\vec{p})|0_\mathcal{D}\rangle$ are naturally interpreted as a 
one-electron states. Here is an array with the properties of the one-electron
states
\begin{align*}
&{H_{\mathcal{D}}}
c^\dagger_s(\vec{p})|0_\mathcal{D}\rangle=E_{\vec{p}}\;\!c^\dagger_s(\vec{p})|0_\mathcal{D}\rangle&
&{H_{\mathcal{D}}}
\zeta^\dagger_s(\vec{p})|0_\mathcal{D}\rangle=-E_{\vec{p}}\;\!\zeta^\dagger_s(\vec{p})|0_\mathcal{D}\rangle&\\
&F
c^\dagger_s(\vec{p})|0_\mathcal{D}\rangle=c^\dagger_s(\vec{p})|0_\mathcal{D}\rangle&
&F
\zeta^\dagger_s(\vec{p})|0_\mathcal{D}\rangle=\zeta^\dagger_s(\vec{p})|0_\mathcal{D}\rangle&\\
&\vec{P}
c^\dagger_s(\vec{p})|0_\mathcal{D}\rangle=\vec{p}\;\!c^\dagger_s(\vec{p})|0_\mathcal{D}\rangle&
&\vec{P}
\zeta^\dagger_s(\vec{p})|0_\mathcal{D}\rangle=\vec{p}\;\!\zeta^\dagger_s(\vec{p})|0_\mathcal{D}\rangle&\\
&\vec{J}
c^\dagger_s(\vec{p})|0_\mathcal{D}\rangle=\vec{p}\;\!c^\dagger_s(\vec{p})|0_\mathcal{D}\rangle&
&\vec{J}
\zeta^\dagger_s(\vec{p})|0_\mathcal{D}\rangle=-\vec{p}\;\!\zeta^\dagger_s(\vec{p})|0_\mathcal{D}\rangle&
\end{align*}
We see that the negative-energy electrons are not appropriated to the
momentum space: it is difficult to give an interpretation of the
state $\zeta^\dagger_s(\vec{p})|0_\mathcal{D}\rangle$ since it has a charge
$-e$, momentum $\vec{p}$ and a charge current $e\vec{p}$.
However, these states are well suited to the study of localized
properties. To see that, let us switch to the Schr\"odinger
picture; the Schr\"odinger fields are the Heisenberg fields taken
at time $t=0$
\begin{equation*}
\psi_S(\vec{x})=\psi(t=0,\vec{x})=\psi(\vec{x})
\end{equation*}
and we have the Schr\"odinger equation 
\begin{equation*}
i\frac{\partial|\Psi(t)\rangle}{\partial t}={H_{\mathcal{D}}}|\Psi(t)\rangle~.
\end{equation*}
The fermion-number $F=\int d^3\vec{x}\psi^\dagger(\vec{x})\psi(\vec{x})$ 
is the total number of positive-energy and negative-energy electrons, and the density of electrons (or fermion-number density) 
is thus given by 
\begin{equation*}
\psi^\dagger(\vec{x})\psi(\vec{x})=\psi^\dagger_a(\vec{x})\psi_a(\vec{x})~.
\end{equation*}
What are the eigenstates of the the fermion-number density? Let us start from a state $|\Phi\rangle$, such that 
\begin{equation}\label{Phi}
\psi^\dagger(\vec{x})\psi(\vec{x})|\Phi\rangle=f(\vec{x})|\Phi\rangle~,
\end{equation}
and apply on operator $\psi^\dagger_b(\vec{y})$ on $|\Phi\rangle$. From eq. (\ref{Phi}) and from the relations (\ref{car}), it can be seen that 
\begin{equation*}
\psi^\dagger(\vec{x})\psi(\vec{x})\psi^\dagger_b(\vec{y})|\Phi\rangle=(\delta(\vec{x}-\vec{y})+f(\vec{x}))\psi^\dagger_b(\vec{y})|\Phi\rangle~,
\end{equation*}
so that any of the four operators $\psi^\dagger_b(\vec{y})$ creates a quantum of the fermion-number at point $\vec{y}$. Thus the signification of 
$\psi^\dagger_b(\vec{y})$ is that it creates an electron at point $\vec{y}$. In the same way, it can be seen that the operators $\psi_b(\vec{y})$ 
are annihilators of electrons at point $\vec{y}$. To obtain the eigenstates of the fermion-number density, one has to start from a state which is 
annihilated by any operator $\psi_b(\vec{y})$ and apply various creators $\psi^\dagger_{a_j}(\vec{x}_j)$ on it. That state is simply the vacuum 
$|0_\mathcal{D}\rangle$:
\begin{equation*}
\psi_b(\vec{x})|0_\mathcal{D}\rangle=0~~\forall\;\!\vec{x}\in\mathbb{R}^3~~\forall\;\!b\in\{1,2,3,4\}~.
\end{equation*}
Here is an array with the first eigenstates of the
fermion-number density.
\begin{center}
\begin{tabular}{|l|l|}
\hline
$F=0$ & $|0_\mathcal{D}\rangle$\\
$F=1$ &
$\psi^\dagger_{a_1}(\vec{x}_1)|0_\mathcal{D}\rangle~\forall~a_1\in\{1,2,3,4\}~\forall\vec{x}_1\in\mathbb{R}^3$\\
$F=2$ &
$\psi^\dagger_{a_1}(\vec{x}_1)\psi^\dagger_{a_2}(\vec{x}_2)|0_\mathcal{D}\rangle~\forall~a_1,a_2\in\{1,2,3,4\}
~\forall\vec{x}_1,\vec{x}_2\in\mathbb{R}^3$\\
\ldots & \ldots \\
\hline
\end{tabular}
\end{center}
The structure of the Fock space is then
\begin{equation*}
\mathcal{F}=\mathcal{H}_1\oplus(\mathcal{H}_1\otimes\mathcal{H}_1)_{as}
\oplus(\mathcal{H}_1\otimes\mathcal{H}_1\otimes\mathcal{H}_1)_{as}\oplus\ldots~.
\end{equation*}
where $\mathcal{H}_1$ is the Hilbert space with fermion-number
number equal to $1$ (thus completely spanned by the orthonormal
basis $\{\psi^\dagger_a(\vec{x})|0_\mathcal{D}\rangle~\forall~a,\vec{x}\}$),
where $\oplus$ is the direct sum, $\otimes$ the direct-product,
and $as$ means that the space is restricted to its antisymmetric
part.

Since $[{H_{\mathcal{D}}},F]=0$ and in accordance with the well-known super-selection rule that
forbids superpositions of states with different values of the
fermion-number, we know that $|\Psi(t)\rangle$ is an eigenstate of
the fermion-number. Let us consider the case where there is only
one quantum of the fermion-number:
\begin{equation*}
F|\Psi(t)\rangle=|\Psi(t)\rangle~.
\end{equation*}
Then $|\Psi(t)\rangle$ can be decomposed along the eigenstates of
the fermion-number density ($\psi^\dagger(\vec{x})\psi(\vec{x})$)
with fermion-number equal to $1$; these eigenstates are
\begin{equation*}
\psi^\dagger_a(\vec{x})|0_\mathcal{D}\rangle~\vec{x}\in\mathbb{R}^3,~~~a\in\{1,2,3,4\}~.
\end{equation*}
Thus, in our case,
\begin{equation*}
|\Psi(t)\rangle=\sum_{a}\int
d^3\vec{x}\Psi_a(t,\vec{x})\psi^\dagger_a(\vec{x})|0_\mathcal{D}\rangle~.
\end{equation*}
Inserting the previous equation in the Schr\"odinger equation,
using the relations (\ref{car}), and the definition of the hamiltonian, one finds
that
\begin{equation*}
i\frac{\partial\Psi(t,\vec{x})}{\partial
t}=-i\vec{\alpha}\cdot\vec{\nabla}\Psi(t,\vec{x})+m\beta\Psi(t,\vec{x})~,
\end{equation*}
which is the Dirac equation. So the link is made between the first
and the second quantization.

Usually, the theory formalism is rewritten by introducing the po\-si\-trons, thanks to the the sub\-sti\-tu\-tions
\begin{align*}
&\zeta_s(\vec{p})\rightarrow d^\dagger_s(-\vec{p})&
&\zeta^\dagger_s(\vec{p})\rightarrow d_s(-\vec{p})~.&
\end{align*}
Then another vacuum has to be defined (let us call it $|0\rangle$):
\begin{align*}
&c_s(\vec{p})|0\rangle=0& &d_s(\vec{p})|0\rangle=0&
&\forall~s,\vec{p}~.&
\end{align*}
That vacuum $|0\rangle$ is the usual vacuum. The
operator $\zeta^\dagger_s(\vec{p})\zeta_s(\vec{p})$ is replaced by
$d_s(-\vec{p})d^\dagger_s(-\vec{p})$, which is also equal to
$-d^\dagger_s(-\vec{p})d_s(-\vec{p})+1$. Making that substitution
in the previous observables leads to
\begin{align*}
&{H_{\mathcal{D}}}=\sum_{s,\vec{p}}\sqrt{|\vec{p}|^2+m^2}(c^\dagger_s(\vec{p})c_s(\vec{p})+d^\dagger_s(\vec{p})d_s(\vec{p}))&\\
&F=\sum_{s,\vec{p}}(c^\dagger_s(\vec{p})c_s(\vec{p})-d^\dagger_s(\vec{p})d_s(\vec{p}))&\\
&\vec{P}=\sum_{s,\vec{p}}\vec{p}\;\!(c^\dagger_s(\vec{p})c_s(\vec{p})+d^\dagger_s(\vec{p})d_s(\vec{p}))&\\
&\vec{J}=\sum_{s,\vec{p}}\vec{p}\;\!(c^\dagger_s(\vec{p})c_s(\vec{p})-d^\dagger_s(\vec{p})d_s(\vec{p}))~,&
\end{align*}
up to some constants which have been dropped. Another
observable is usually defined, when we talk of positrons and
electrons; it is the particle-number
\begin{equation*}
N=\sum_{s,\vec{p}}(c^\dagger_s(\vec{p})c_s(\vec{p})+d^\dagger_s(\vec{p})d_s(\vec{p}))~.
\end{equation*}
The properties of the one-particle states are
\begin{align*}
&{H_{\mathcal{D}}}
c^\dagger_s(\vec{p})|0\rangle=\sqrt{|\vec{p}|^2+m^2}c^\dagger_s(\vec{p})|0\rangle&
&{H_{\mathcal{D}}}
d^\dagger_s(\vec{p})|0\rangle=\sqrt{|\vec{p}|^2+m^2}d^\dagger_s(\vec{p})|0\rangle&\\
&F
c^\dagger_s(\vec{p})|0\rangle=c^\dagger_s(\vec{p})|0\rangle&
&F
d^\dagger_s(\vec{p})|0\rangle=-d^\dagger_s(\vec{p})|0\rangle&\\
&\vec{P}
c^\dagger_s(\vec{p})|0\rangle=\vec{p}\;\!c^\dagger_s(\vec{p})|0\rangle&
&\vec{P}
d^\dagger_s(\vec{p})|0\rangle=\vec{p}\;\!d^\dagger_s(\vec{p})|0\rangle&\\
&\vec{J}
c^\dagger_s(\vec{p})|0\rangle=\vec{p}\;\!c^\dagger_s(\vec{p})|0\rangle&
&\vec{J}
d^\dagger_s(\vec{p})|0\rangle=-\vec{p}\;\!d^\dagger_s(\vec{p})|0\rangle~.&
\end{align*}
Thus, when we study properties related to the momentum space, the
electrons and positrons point of view is the best one. However, the localized-particle states 
are not so natural. These states are defined as the eigenstates of the operator
\begin{equation}\label{partdens}
n(\vec{x})=\sum_{s=1}^{s=2}[C^\dagger_s(\vec{x})C_s(\vec{x})+D^\dagger_s(\vec{x})D_s(\vec{x})]~,
\end{equation}
where 
\begin{align*}
&C_s(\vec{x})=\frac{1}{\sqrt{(2\pi)^3}}\int d^3\vec{p} c_s(\vec{p})e^{i\vec{p}\cdot\vec{x}}& 
&D_s(\vec{x})=\frac{1}{\sqrt{(2\pi)^3}}\int d^3\vec{p} d_s(\vec{p})e^{i\vec{p}\cdot\vec{x}}~.&
\end{align*}
Eigenstates of the particle density are obtained from the vacuum $|0\rangle$ by applying several creators $C^\dagger_{s_j}(\vec{x}_j)$ or 
$D^\dagger_{s_k}(\vec{x}_k)$ on it. These states are the Newton-Wigner states.

The particle-number is different from the fermion-number
\begin{equation*}
F=\sum_{s,\vec{p}}(c^\dagger_s(\vec{p})c_s(\vec{p})+d_s(\vec{p})d^\dagger_s(\vec{p}))=
C+\sum_{s,\vec{p}}(c^\dagger_s(\vec{p})c_s(\vec{p})-d^\dagger_s(\vec{p})d_s(\vec{p}))
\end{equation*}
(where $C$ is an infinite constant). It is worth
mentioning that the particle-number does not commute with the
fermion-number density $\psi^\dagger(\vec{x})\psi(\vec{x})$. It is
possible to find well-behaved functions $f$ such that
\begin{equation*}
[\int d^3\vec{x}
f(\vec{x})\psi^\dagger(\vec{x})\psi(\vec{x}),N]\neq 0~.
\end{equation*}
The proof is given in appendix \ref{sec:appb}. In
the electrons and positrons point of view, the charge density is
defined as
\begin{equation*}
Q(\vec{x})=-e:\psi^\dagger(\vec{x})\psi(\vec{x}):~,
\end{equation*}
where the dots mean normal-ordering. A corollary is that
eigenstates of the charge density are not eigenstates of the
particle-number.
\section{Localized measurements}
In the non-relativistic pilot-wave theory, the hidden variables are the positions of the particles. In the quantum field theory context, what hidden variables 
must we choose? The first natural answer seems to be the particle density. We will now argue why the particle-density (eq. (\ref{partdens})) 
is not a good choice for a beable.  

The first point is that the observable $n(\vec{x})$ appears very artificial, since it is not made of the fields $\psi(\vec{x})$ and
$\psi^\dagger(\vec{x})$. Remember that any measurement is made with a measuring apparatus that correlate eigenstates of the observable being 
measured to eigenstates of the apparatus. So, when we look at localized properties, we are interested in hamiltonian densities. 
Of course, $n(\vec{x})$ is hermitian, but that does not imply that it has 
a physical meaning. In fact, when we measure localized properties, if we want to attain high precision, 
we have to use high energy (small wavelengths), and that leads to pairs creation. So it seems that the idea of an eigenstate containing 
a particle localized in a small region, looses his meaning when the region 
is sufficiently small. There is a second argument that shows that $n(\vec{x})$ does not have physical meaning: if we start from a state 
$C^\dagger_s(\vec{x})|0\rangle$ at time $t$ and if we let it evolve according to the Schr\"odinger equation for a small time $\Delta t$, then it 
can be shown that it has a small probability to be outside the light-cone of the event $(\vec{x},t)$ (see \cite{hegerfeldt} and 
also \cite{teller}). The third point is 
more related to the general ideas of the pilot-wave theory. In fact, the crucial ingredient of the pilot-wave theory is the existence of 
a continuity equation linked to the chosen beable. Since the particle number is not conserved, we do not expect such an equation for the 
particle density.

In the non-relativistic pilot-wave theory, the hidden variable is the particle density, but since there are only electrons, the particle density 
is proportional to the charge density. Then can we take the charge density as the beable? That has been suggested by Bell, 
in his interpretation of the lattice fermionic quantum field theories \cite{bell1}. To be more precise, Bell suggested that 
the fermion-number density $\psi^\dagger(\vec{x})\psi(\vec{x})$ could be given the beable status. All the critics that we made 
of the particle density do not apply to the fermion-number density. The fermion-number density does not commute with the 
particle number. If we start from a state $\psi^\dagger_a(\vec{x})|0_{\mathcal{D}}\rangle$ at time $t$, the charge will stay in the 
light cone. And finally, the fermion-number is conserved, so we expect a continuity equation.

\section{\label{sec:dbm}The Bell model for the free Dirac quantum field theory}
Since the fermion-number is conserved ($[{H_{\mathcal{D}}},F]=0$), and in accordance with the well-known super-selection rule, 
the pilot-state $|\Psi(t)\rangle$ is an eigenstate of the fermion-number; let us define the corresponding eigenvalue 
by $\omega$: $F|\Psi(t)\rangle=\omega|\Psi(t)\rangle$. 
So the pilot-state can be decomposed along the eigenstates of the fermion-number density with fermion-number equal to $\omega$:
\begin{equation*}
|\Psi(t)\rangle=\frac{1}{\omega!}\int
d^3\vec{x}_1\cdot\cdot d^3\vec{x}_\omega\Psi_{a_1\cdot\cdot
a_\omega}(t,\vec{x}_1,\cdot\cdot,\vec{x}_\omega)
\psi^\dagger_{a_1}(\vec{x}_1)\cdot\cdot\psi^\dagger_{a_\omega}(\vec{x}_\omega)
|0_\mathcal{D}\rangle~
\end{equation*}
(from now on, we use Einstein's convention: any repeated index is summed over).
It is thus possible, from the Schr\"odinger equation, to obtain an equation satisfied by $\Psi_{a_1\cdot\cdot
a_\omega}(t,\vec{x}_1,\cdot\cdot,\vec{x}_\omega)$. It can be shown that this equation is 
\begin{equation}\label{superse}
i\frac{\partial\Psi_{a_1\cdot\cdot
a_\omega}(t,\vec{X})}{\partial t}=\sum_{j=1}^{j=\omega}
[-i\vec{\alpha}_{a_j a}\cdot\vec{\nabla}_{\vec{x}_j}+m\beta_{a_j a}]\Psi_{a_1\cdot\cdot(a_j\rightarrow a)\cdot\cdot
a_\omega}(t,\vec{X})~,
\end{equation}
where $\vec{X}$ is a short notation for $(\vec{x}_1,\ldots,\vec{x}_\omega)$.
We consider the case $\omega=2$. It can be easily generalized. The pilot-state is 
\begin{equation*}
|\Psi(t)\rangle=\frac{1}{2!}\int d^3\vec{x}_1d^3\vec{x}_2\Psi_{a_1 a_2}(t,\vec{x}_1,\vec{x}_2)
\psi^\dagger_{a_1}(\vec{x}_1)\psi^\dagger_{a_2}(\vec{x}_2)|0_\mathcal{D}\rangle~.
\end{equation*}
Let us apply the hamiltonian on $|\Psi(t)\rangle$:
\begin{align}\label{greq}
{H_{\mathcal{D}}}&|\Psi(t)\rangle=\frac{1}{2!}\int d^3\vec{x}_1 d^3\vec{x}_2 d^3\vec{x}\Psi_{a_1 a_2}(t,\vec{x}_1,\vec{x}_2)&\nonumber\\
&[i\vec{\nabla}\psi^\dagger_a(\vec{x})\cdot\vec{\alpha}_{ab}+m\psi^\dagger_a(\vec{x})\beta_{a b}]\psi_b(\vec{x})
\psi^\dagger_{a_1}(\vec{x}_1)\psi^\dagger_{a_2}(\vec{x}_2)|0_\mathcal{D}\rangle&\nonumber\\
=&\frac{1}{2!}\int d^3\vec{x}_1 d^3\vec{x}_2 d^3\vec{x}\Psi_{a_1 a_2}(t,\vec{x}_1,\vec{x}_2)
[i\vec{\nabla}\psi^\dagger_a(\vec{x})\cdot\vec{\alpha}_{ab}+m\psi^\dagger_a(\vec{x})\beta_{a b}]&\nonumber\\
&(\delta_{a_1 b}\delta(\vec{x}-\vec{x}_1)\psi^\dagger_{a_2}(\vec{x}_2)-\delta_{a_2 b}\delta(\vec{x}-\vec{x}_2)
\psi^\dagger_{a_1}(\vec{x}_1))|0_\mathcal{D}\rangle&\nonumber\\
=&\frac{1}{2!}\int d^3\vec{x}_1 d^3\vec{x}_2 \Psi_{a_1 a_2}(t,\vec{x}_1,\vec{x}_2)&\nonumber\\
[&i\vec{\alpha}_{a a_1}\cdot\vec{\nabla}\psi^\dagger_a(\vec{x_1})
\psi^\dagger_{a_2}(\vec{x}_2)|0_\mathcal{D}\rangle-i\vec{\alpha}_{a a_2}\cdot\vec{\nabla}\psi^\dagger_a(\vec{x_2})
\psi^\dagger_{a_1}(\vec{x}_1)|0_\mathcal{D}\rangle&\nonumber\\
+&m\beta_{a a_1}\psi^\dagger_a(\vec{x_1})
\psi^\dagger_{a_2}(\vec{x}_2)|0_\mathcal{D}\rangle-m\beta_{a a_2}\psi^\dagger_a(\vec{x_2})
\psi^\dagger_{a_1}(\vec{x}_1)|0_\mathcal{D}\rangle]&\nonumber\\
=&\frac{1}{2!}\int d^3\vec{x}_1 d^3\vec{x}_2 \Psi_{a_1 a_2}(t,\vec{x}_1,\vec{x}_2)&\nonumber\\
[&i\vec{\alpha}_{a a_1}\cdot\vec{\nabla}\psi^\dagger_a(\vec{x_1})
\psi^\dagger_{a_2}(\vec{x}_2)|0_\mathcal{D}\rangle+i\vec{\alpha}_{a a_2}\cdot
\psi^\dagger_{a_1}(\vec{x}_1)\vec{\nabla}\psi^\dagger_a(\vec{x_2})|0_\mathcal{D}\rangle&\nonumber\\
+&m\beta_{a a_1}\psi^\dagger_a(\vec{x_1})
\psi^\dagger_{a_2}(\vec{x}_2)|0_\mathcal{D}\rangle+m\beta_{a a_2}\psi^\dagger_{a_1}(\vec{x}_1)\psi^\dagger_a(\vec{x_2})|0_\mathcal{D}\rangle]~.&
\end{align}
The second equality comes from the relations (\ref{car}) and from the fact that $|0_\mathcal{D}\rangle$ is annihilated by any operator 
$\psi_a(\vec{x})$. The third equality results from an integration over $\vec{x}$. The fourth equality is a consequence of the anti-commutation 
of the fields $\psi^\dagger$. Integrating by parts eq. (\ref{greq}) and renaming the dummy indices gives
\begin{align*}
{H_{\mathcal{D}}}|\Psi(t)\rangle=&\frac{1}{2!}\int d^3\vec{x}_1 d^3\vec{x}_2
[-i\vec{\alpha}_{a_1 a}\cdot\vec{\nabla}_{\vec{x}_1}\Psi_{a a_2}(t,\vec{x}_1,\vec{x}_2)&\\
-&i\vec{\alpha}_{a_2 a}\cdot\vec{\nabla}_{\vec{x}_2}\Psi_{a_1 a}(t,\vec{x}_1,\vec{x}_2)
+m\beta_{a_1 a}\Psi_{a a_2}(t,\vec{x}_1,\vec{x}_2)&\\+&m\beta_{a_2 a}\Psi_{a_1 a}(t,\vec{x}_1,\vec{x}_2)]\psi^\dagger_a(\vec{x_1})
\psi^\dagger_{a_2}(\vec{x}_2)|0_\mathcal{D}\rangle~.&
\end{align*}
And since 
\begin{equation*}
i\frac{\partial \Psi(t)\rangle}{\partial t}=\frac{i}{2!}\int d^3\vec{x}_1d^3\vec{x}_2\frac{\partial\Psi_{a_1 a_2}(t,\vec{x}_1,\vec{x}_2)}{\partial t}
\psi^\dagger_{a_1}(\vec{x}_1)\psi^\dagger_{a_2}(\vec{x}_2)|0_\mathcal{D}\rangle~,
\end{equation*}
the Schr\"odinger equation is equivalent to the above-mentioned result (eq. (\ref{superse})) for the case $\omega=2$.

The probability density to observe the universe in a configuration $(\vec{x}_1,\ldots,\vec{x}_\omega)$ is 
\begin{equation}\label{pde}
\rho(t,\vec{x}_1,\ldots,\vec{x}_\omega)=
\Psi^*_{a_1\cdot\cdot
a_\omega}(t,\vec{x}_1,\cdot\cdot,\vec{x}_\omega)\Psi_{a_1\cdot\cdot
a_\omega}(t,\vec{x}_1,\cdot\cdot,\vec{x}_\omega)~.
\end{equation}
With the help of equation (\ref{superse}), it can be shown that we have the following 
continuity equation
\begin{equation}\label{cebell}
\frac{\partial\rho(t,\vec{X})}{\partial t}+\sum_{k=1}^{k=\omega}
\vec{\nabla}_{\vec{x}_k}\cdot\vec{j}_k(t,\vec{X})=0~,
\end{equation}
where 
\begin{equation*}
\vec{j}_k(t,\vec{x}_1,\ldots,\vec{x}_\omega)=
\Psi^*_{a_1\cdot\cdot
a_\omega}(t,\vec{x}_1,\cdot\cdot,\vec{x}_\omega)\vec{\alpha}_{a_j a}\Psi_{a_1\cdot\cdot(a_j\rightarrow a)\cdot\cdot
a_\omega}(t,\vec{x}_1,\cdot\cdot\vec{x}_\omega)~.
\end{equation*}
Let us go on with the proof: with the help of eq. (\ref{superse}), the time-derivative of $\rho(t,\vec{X})$ (eq. (\ref{pde})) is equal to
\begin{align*}
\dot{\rho}(t,\vec{X})=-&\Psi^{*}_{a_1\cdot\cdot
a_\omega}(t,\vec{X})\sum_{j=1}^{j=\omega}
[\vec{\alpha}_{a_j a}\cdot\vec{\nabla}_{\vec{x}_j}+im\beta_{a_j a}]\Psi_{a_1\cdot\cdot(a_j\rightarrow a)\cdot\cdot
a_\omega}(t,\vec{X})&\\
-&\sum_{j=1}^{j=\omega}\vec{\nabla}_{\vec{x}_j}\Psi^*_{a_1\cdot\cdot(a_j\rightarrow a)\cdot\cdot
a_\omega}(t,\vec{X})\cdot(\vec{\alpha}_{a_j a})^{*}\Psi_{a_1\cdot\cdot
a_\omega}(t,\vec{X})&\\
+&im\sum_{j=1}^{j=\omega}\Psi^*_{a_1\cdot\cdot(a_j\rightarrow a)\cdot\cdot
a_\omega}(t,\vec{X})(\beta_{a_j a})^{*}\Psi_{a_1\cdot\cdot a_\omega}(t,\vec{X})&
\end{align*}
Noting that $(\vec{\alpha}_{a_j a})^{*}=(\vec{\alpha}^\dagger)_{a a_j}=\vec{\alpha}_{a a_j}$, that  
$(\beta_{a_j a})^{*}=(\beta^\dagger)_{a a_j}=\beta_{a a_j}$, and renaming the dummy indices, we see that the terms containing a factor $m$ 
cancel each other out. So there remains 
\begin{align*}
\dot{\rho}(t,\vec{X})=-&\Psi^{*}_{a_1\cdot\cdot
a_\omega}(t,\vec{X})\sum_{j=1}^{j=\omega}
[\vec{\alpha}_{a_j a}\cdot\vec{\nabla}_{\vec{x}_j}]\Psi_{a_1\cdot\cdot(a_j\rightarrow a)\cdot\cdot
a_\omega}(t,\vec{X})&\\
-&\sum_{j=1}^{j=\omega}\vec{\nabla}_{\vec{x}_j}\Psi^*_{a_1\cdot\cdot(a_j\rightarrow a)\cdot\cdot
a_\omega}(t,\vec{X})\cdot\vec{\alpha}_{a a_j}\Psi_{a_1\cdot\cdot
a_\omega}(t,\vec{X})~,
\end{align*}
and the conclusion (eq. (\ref{cebell})) follows directly. 

All this expressions are totally similar to those found in the non-relativistic pilot-wave theory. The positions of the fermions (electrons of 
positive and negative energy) play the role of the positions of the particles in the non-relativistic theory. Their number is conserved in both cases; that 
leads to continuity equations, eq. (\ref{cebell}) being the analog of eq. (\ref{ce}). In the Bell model of the free Dirac 
quantum field theory, the universe is described by a pilot-state, solution of the Schr\"odinger equation, and by a point 
$(\vec{x}_1(t),\ldots,\vec{x}_\omega(t))$. Proceeding just as in the non-relativistic pilot-wave theory (see eq. (\ref{vlaw})), it is easy to see that 
if the kth fermion is moving according to the velocity-law 
\begin{equation*}
\vec{v}_k(t)=\frac{\Psi^*_{a_1\cdot\cdot
a_\omega}(t,\vec{x}_1,\cdot\cdot,\vec{x}_\omega)\vec{\alpha}_{a_k a}\Psi_{a_1\cdot\cdot(a_k\rightarrow a)\cdot\cdot
a_\omega}(t,\vec{x}_1,\cdot\cdot\vec{x}_\omega)}{\Psi^*_{b_1\cdot\cdot
b_\omega}(t,\vec{x}_1,\cdot\cdot,\vec{x}_\omega)\Psi_{b_1\cdot\cdot
b_\omega}(t,\vec{x}_1,\cdot\cdot,\vec{x}_\omega)}~,
\end{equation*}
then all the predictions of the orthodox quantum field theory are regained. Of course, for any physical state (any state obtained 
from $|0\rangle$ by creating a finite number of electrons and positrons), $\omega$ is infinite. But the number of fermions $\omega$ is 
countable if the volume of the universe is finite. Another remark is that is also totally equivalent to say that the charge-density is the element of 
reality.
\section{A deterministic Bell model for quantum electrodynamics}
The complete hamiltonian for quantum electrodynamics is 
\begin{equation*}
H=H_{\mathcal{D}}+H_{\Gamma}-e\int d^3\vec{x}\bar{\psi}(\vec{x})\gamma^\mu\psi(\vec{x})A_\mu(\vec{x})~,
\end{equation*}
where $H_{\mathcal{D}}$ is the free Dirac hamiltonian and $H_{\Gamma}$ is the free photon hamiltonian. The pilot-state is defined as the solution of the 
Schr\"odinger equation 
\begin{equation*}
i\frac{\partial|\Psi(t)\rangle}{\partial t}=H|\Psi(t)\rangle~.
\end{equation*}
Since $[H,F]=0$, and according to the charge super selection rule, the pilot-state is an eigenstate of the fermion-number; let us define the corresponding 
eigenvalue by $\omega$
\begin{equation*}
F|\Psi(t)\rangle=\omega|\Psi(t)\rangle~.
\end{equation*}
A complete basis for that particular class of states is then given by 
\begin{align*}
&|\gamma\rangle\otimes\psi^\dagger_{a_1}(\vec{x}_1)\ldots\psi^\dagger_{a_\omega}(\vec{x}_\omega)|0_{\mathcal{D}}\rangle&\\
&\vec{x}_1,\ldots,\vec{x}_\omega\in \mathbb{R}^3~~a_1,\ldots,a_\omega\in\{1,2,3,4\}~~\gamma\in\{\gamma\}~,&
\end{align*}
where the set of all $|\gamma\rangle$ form a complete basis of the photons Fock space and where 
$\psi^\dagger_{a_1}(\vec{x}_1)\ldots\psi^\dagger_{a_\omega}(\vec{x}_\omega)|0_{\mathcal{D}}\rangle$ is an eigenstate 
of the fermion density with fermion-number equal to $\omega$. Hence we have that 
\begin{align*}
|\Psi(t)\rangle=\sum_{a_1=1}^{a_1=4}\cdot\cdot\sum_{a_\omega=1}^{a_\omega=4}\sum_{\gamma}\int d^3\vec{x}_1\cdot\cdot d^3\vec{x}_\omega
&\Psi^\gamma_{a_1\ldots a_\omega}(t,\vec{x}_1,\ldots,\vec{x}_\omega)&\\
&|\gamma\rangle\otimes\psi^\dagger_{a_1}(\vec{x}_1)\ldots\psi^\dagger_{a_\omega}(\vec{x}_\omega)|0_{\mathcal{D}}\rangle~.&
\end{align*}
Inserting that expression in the Schr\"odinger equation, it is possible to show that the equation satisfied by 
$\Psi^\gamma_{a_1\ldots a_\omega}(t,\vec{x}_1,\ldots,\vec{x}_\omega)$ is 
\begin{align}\label{superseqed} i\frac{\partial\Psi^{\gamma}_{a_1\cdot\cdot a_\omega}(t,\vec{X})}{\partial t}=
&\sum_{j=1}^{j=\omega}[-i\vec{\alpha}_{a_j a}\cdot\vec{\nabla}_{\vec{x}_j}+m\beta_{a_j a}]\Psi^{\gamma}_{a_1\cdot\cdot(a_j\rightarrow a)\cdot\cdot
a_\omega}(t,\vec{X})&\nonumber\\ -&\sum_{j=1}^{j=\omega}\sum_{\gamma'}
\langle\gamma|e(\beta\gamma^\mu)_{a_j a}A_\mu(\vec{x}_j)|\gamma'\rangle\Psi^{\gamma'}_{a_1\cdot\cdot(a_j\rightarrow a)\cdot\cdot
a_\omega}(t,\vec{X})\nonumber&\\ +
&\sum_{\gamma'}\langle\gamma|H_{\Gamma}|\gamma'\rangle\Psi^{\gamma'}_{a_1\cdot\cdot
a_\omega}(t,\vec{X})~.&
\end{align}
The probability density to observe the system in a configuration $\vec{X}$ at time $t$ is 
\begin{equation*}
\rho(t,\vec{X})=\Psi^{\gamma *}_{a_1\cdot\cdot
a_\omega}(t,\vec{X})\Psi^\gamma_{a_1\cdot\cdot\cdot
a_\omega}(t,\vec{X})
\end{equation*}
(sum over any repeated index). Let us calculate $\dot{\rho}(t,\vec{X})$ with the help of eq. (\ref{superseqed}). We have that $\dot{\rho}(t,\vec{X})=$
\begin{align*}
&-\Psi^{\gamma *}_{a_1\cdot\cdot a_\omega}(t,\vec{X})\sum_{j=1}^{j=\omega}
[\vec{\alpha}_{a_j a}\cdot\vec{\nabla}_{\vec{x}_j}+im\beta_{a_j a}]\Psi^{\gamma}_{a_1\cdot\cdot(a_j\rightarrow a)\cdot\cdot
a_\omega}(t,\vec{X})&\\
&-\sum_{j=1}^{j=\omega}\vec{\nabla}_{\vec{x}_j}\Psi^{\gamma *}_{a_1\cdot\cdot(a_j\rightarrow a)\cdot\cdot
a_\omega}(t,\vec{X})\cdot(\vec{\alpha}_{a_j a})^{*}\Psi^{\gamma}_{a_1\cdot\cdot
a_\omega}(t,\vec{X})&\\
&+im\sum_{j=1}^{j=\omega}\Psi^{\gamma *}_{a_1\cdot\cdot(a_j\rightarrow a)\cdot\cdot
a_\omega}(t,\vec{X})(\beta_{a_j a})^{*}\Psi^{\gamma}_{a_1\cdot\cdot a_\omega}(t,\vec{X})&\\
&+i\sum_{\gamma\;\!\gamma'}\Psi^{\gamma *}_{a_1\cdot\cdot
a_\omega}(t,\vec{X})\sum_{j=1}^{j=\omega}
\langle\gamma|e(\beta\gamma^\mu)_{a_j a}A_\mu(\vec{x}_j)|\gamma'\rangle\Psi^{\gamma'}_{a_1\cdot\cdot(a_j\rightarrow a)\cdot\cdot
a_\omega}(t,\vec{X})&
\end{align*}
\begin{align*}
&-i\sum_{\gamma\;\!\gamma'}\Psi^{\gamma *}_{a_1\cdot\cdot a_\omega}(t,\vec{X})\langle\gamma|H_{\Gamma}|\gamma'\rangle
\Psi^{\gamma'}_{a_1\cdot\cdot(a_j\rightarrow a)\cdot\cdot a_\omega}(t,\vec{X})&\\
&-i\sum_{\gamma\;\!\gamma'}\sum_{j=1}^{j=\omega}\Psi^{\gamma' *}_{a_1\cdot\cdot(a_j\rightarrow a)\cdot\cdot
a_\omega}(t,\vec{X})
\langle\gamma'|e(\beta\gamma^\mu)^*_{a a_j} A^\dagger_\mu(\vec{x}_j)|\gamma\rangle\Psi^{\gamma}_{a_1\cdot\cdot a_\omega}(t,\vec{X})&\\
&+i\sum_{\gamma\;\!\gamma'}\Psi^{\gamma' *}_{a_1\cdot\cdot(a_j\rightarrow a)\cdot\cdot
a_\omega}(t,\vec{X})\langle\gamma'|H^\dagger_{\gamma}|\gamma
\rangle\Psi^{\gamma}_{a_1\cdot\cdot a_\omega}(t,\vec{X})~.&
\end{align*}
Noting that the matrix $\beta$ is hermitian and that the dummy indices can be renamed, it is easy to check that 
the terms containing a factor $m$ cancel each other out. In the same way, on account of the hermiticity of $H_{\Gamma}$, 
$A^\dagger_\mu(\vec{x})$ and $(\beta\gamma^\mu)$, and due to the possible renaming of the dummy indices, 
the terms containing photon fields cancel each other out, so that  
\begin{eqnarray*}\dot{\rho}(t,\vec{X})=
-\Psi^{\gamma *}_{a_1\cdot\cdot
a_\omega}(t,\vec{X})\sum_{j=1}^{j=\omega}
[\vec{\alpha}_{a_j a}\cdot\vec{\nabla}_{\vec{x}_j}]\Psi^{\gamma}_{a_1\cdot\cdot(a_j\rightarrow a)\cdot\cdot
a_\omega}(t,\vec{X})\\
-\sum_{j=1}^{j=\omega}\vec{\nabla}_{\vec{x}_j}\Psi^{\gamma *}_{a_1\cdot\cdot(a_j\rightarrow a)\cdot\cdot
a_\omega}(t,\vec{X})\cdot\vec{\alpha}_{a a_j}\Psi^{\gamma}_{a_1\cdot\cdot
a_\omega}(t,\vec{X})~.
\end{eqnarray*}
That last equation can be rewritten as 
\begin{equation*}
\frac{\partial\rho(t,\vec{X})}{\partial t}+\sum_{k=1}^{k=\omega}
\vec{\nabla}_{\vec{x}_k}\cdot\vec{j}_k(t,\vec{X})=0~,
\end{equation*}
where 
\begin{equation*}\vec{j}_k(t,\vec{x}_1,\ldots,\vec{x}_\omega)
=\Psi^{\gamma *}_{a_1\cdot\cdot
a_\omega}(t,\vec{x}_1,\cdot\cdot,\vec{x}_\omega)\vec{\alpha}_{a_j a}\Psi^{\gamma}_{a_1\cdot\cdot(a_j\rightarrow a)\cdot\cdot
a_\omega}(t,\vec{x}_1,\cdot\cdot\vec{x}_\omega)~.
\end{equation*} 
Then, if the kth fermion moves with velocity 
\begin{equation*}
\vec{v}_k(t)=\frac{\Psi^{\gamma *}_{a_1\cdot\cdot
a_\omega}(t,\vec{X})\vec{\alpha}_{a_k a}\Psi^\gamma_{a_1\cdot\cdot(a_k\rightarrow a)\cdot\cdot
a_\omega}(t,\vec{X})}{\Psi^{\gamma *}_{b_1\cdot\cdot
b_\omega}(t,\vec{X})\Psi^\gamma_{b_1\cdot\cdot
b_\omega}(t,\vec{X})}\bigg|_{\vec{X}=\vec{X}(t)}~,
\end{equation*}
all the predictions of the orthodox interpretation of quantum electrodynamics are regained, if any measurement amounts to a measurement 
of the charge density. 
\section{Conclusion and open questions}
We hope to have shown that it is possible to have a picture of what is 
going on in quantum electrodynamics, by taking the fermion-number density, or charge density, as an element 
of reality. Of course, this is still a work in progress. The point of view we have adopted is not 
convenient for practical calculations, but that was not the goal we were pursuing.

A line of research that should be followed is the study of the non-relativistic limit of the Bell model for quantum 
electrodynamics, in order to show if we obtain the non-relativistic pilot-wave theory in that limit. Another line of research 
is the application of these ideas to bosonic quantum field theories. On one hand, in the Klein-Gordon quantum field theory, there 
is no state annihilated by a charge annihilator, so it is an argument against the Bell model for the Klein-Gordon quantum field theory, 
at least in the present form of the Bell model. On the other hand, charge is conserved, so we expect a continuity equation 
for the charge. Perhaps that the solution will come by looking at things locally. 

\paragraph{Thanks}
I would like to thank Thomas Durt, for his many useful comments and his critical reading of the text. I would like to 
thank Thomas Durt and Jacques Robert, for allowing me to participate at the Peyresq congress on electromagnetism, 
and enjoy such an open-minded and friendly 
atmosphere. I would like to thank my advisor, Jean Bricmont.
\newpage
\appendix
\section{\label{sec:appb}Calculation of the commutator $[N,\psi^\dagger(\vec{x})\psi(\vec{x})]$}
We want to show that
\begin{equation}
[N,\psi^\dagger(\vec{x})\psi(\vec{x})]\neq 0~,
\end{equation}
where
\begin{equation}
N=\sum_{r}\int{d^3\vec{k}}[c^\dagger_r(\vec{k})c_r(\vec{k})+d^\dagger_r(\vec{k})d_r(\vec{k})]~,
\end{equation}
even we think about the fields as distributions.
We use the following relation ($F$ stands for fermion):$[F_1F_2,F_3F_4]$
\begin{align*}
=&F_1[F_2,F_3F_4]+[F_1,F_3F_4]F_2&\\
=&F_1\{F_2,F_3\}F_4-F_1F_3\{F_2,F_4\}+\{F_1,F_3\}F_4F_2-F_3\{F_1,F_4\}F_2~.&
\end{align*}
Let us recall the expressions of the spinor fields:
\begin{align*}
&\psi(\vec{x})=\sqrt{\frac{1}{(2\pi)^3}}\sum_{s}\int
{d^3\vec{p}}\sqrt{\frac{m}{E_{\vec{p}}}}[
u_s(\vec{p})e^{i\vec{p}\cdot\vec{x}}c_s(\vec{p})+
v_s(\vec{p})e^{-i\vec{p}\cdot\vec{x}}d^\dagger_s(\vec{p})]&\\
&\psi^\dagger(\vec{x})=\sqrt{\frac{1}{(2\pi)^3}}\sum_{s}\int
{d^3\vec{p}}\sqrt{\frac{m}{E_{\vec{p}}}}[
u^\dagger_s(\vec{p})e^{-i\vec{p}\cdot\vec{x}}c^\dagger_s(\vec{p})+
v^\dagger_s(\vec{p})e^{i\vec{p}\cdot\vec{x}}d_s(\vec{p})]~.&
\end{align*}
By using the anti-commutation relations
\begin{align*}
&\{c_s(\vec{k}),c^\dagger_r(\vec{p})\}=\delta_{sr}\delta^3(\vec{k}-\vec{p})&
&\{d_s(\vec{k}),d^\dagger_r(\vec{p})\}=\delta_{sr}\delta^3(\vec{k}-\vec{p})~,&
\end{align*}
and all other anti-commutators vanishing, we find that
\begin{align*}
&\{\psi^\dagger_a(\vec{x}),c_r(\vec{k})\}=\sqrt{\frac{1}{(2\pi)^3}}
\sqrt{\frac{m}{E_{\vec{k}}}}u^\dagger_{a
r}(\vec{k})e^{-i\vec{k}\cdot\vec{x}}&
&\{\psi_a(\vec{x}),c_r(\vec{k})\}=0&\\
&\{\psi_a(\vec{x}),c^\dagger_r(\vec{k})\}=\sqrt{\frac{1}{(2\pi)^3}}
\sqrt{\frac{m}{E_{\vec{k}}}}u_{a
r}(\vec{k})e^{i\vec{k}\cdot\vec{x}}&
&\{\psi^\dagger_a(\vec{x}),c^\dagger_r(\vec{k})\}=0&\\
&\{\psi_a(\vec{x}),d_r(\vec{k})\}=\sqrt{\frac{1}{(2\pi)^3}}
\sqrt{\frac{m}{E_{\vec{k}}}}v_{a
r}(\vec{k})e^{-i\vec{k}\cdot\vec{x}}&
&\{\psi^\dagger_a(\vec{x}),d_r(\vec{k})\}=0&\\
&\{\psi^\dagger_a(\vec{x}),d^\dagger_r(\vec{k})\}=\sqrt{\frac{1}{(2\pi)^3}}
\sqrt{\frac{m}{E_{\vec{k}}}}v^\dagger_{a
r}(\vec{k})e^{i\vec{k}\cdot\vec{x}}&
&\{\psi_a(\vec{x}),d^\dagger_r(\vec{k})\}=0~,&
\end{align*}
so that $[\psi^\dagger_a(\vec{x})\psi_a(\vec{x}),\sum_{r}\int{d^3\vec{k}}c^\dagger_r(\vec{k})c_r(\vec{k})]$
\begin{align*}
=&\sum_{r}\int{d^3\vec{k}}\bigl(\psi^\dagger_a(\vec{x})\{\psi_a(\vec{x}),c^\dagger_r(\vec{k})\}
c_r(\vec{k})-c^\dagger_r(\vec{k})\{\psi^\dagger_a(\vec{x}),c_r(\vec{k})\}
\psi_a(\vec{x})\bigr)&\\=&\frac{m^2}{(2\pi)^3}\sum_{s,r}\int
\frac{{d^3\vec{p}}{d^3\vec{k}}}{\sqrt{E_{\vec{p}}E_{\vec{k}}}}[
u^\dagger_s(\vec{p})u_r(\vec{k})e^{-i(\vec{p}-\vec{k})\cdot\vec{x}}c^\dagger_s(\vec{p})c_r(\vec{k})&\\&+
v^\dagger_s(\vec{p})u_r(\vec{k})e^{i(\vec{p}+\vec{k})\cdot\vec{x}}d_s(\vec{p})c_r(\vec{k})
-u^\dagger_r(\vec{k})u_s(\vec{p})e^{i(\vec{p}-\vec{k})\cdot\vec{x}}
c^\dagger_r(\vec{k})c_s(\vec{p})&\\&
-u^\dagger_r(\vec{k})v_s(\vec{p})e^{-i(\vec{p}+\vec{k})\cdot\vec{x}}
c^\dagger_r(\vec{k})d^\dagger_s(\vec{p})]~.&
\end{align*}
Since $r$, $s$, $\vec{p}$ and $\vec{k}$ are dummy variables, we
find that
\begin{align*}
&[\psi^\dagger_a(\vec{x})\psi_a(\vec{x}),\sum_{r}\int{d^3\vec{k}}c^\dagger_r(\vec{k})c_r(\vec{k})]=\frac{m^2}{(2\pi)^3}\sum_{s,r}&\\
&\int
\frac{{d^3\vec{p}}{d^3\vec{k}}}{\sqrt{E_{\vec{p}}E_{\vec{k}}}}[
v^\dagger_s(\vec{p})u_r(\vec{k})e^{i(\vec{p}+\vec{k})\cdot\vec{x}}d_s(\vec{p})c_r(\vec{k})]-
&\\
&\frac{m^2}{(2\pi)^3}\sum_{s,r}\int
\frac{{d^3\vec{p}}{d^3\vec{k}}}{\sqrt{E_{\vec{p}}E_{\vec{k}}}}[
u^\dagger_r(\vec{k})v_s(\vec{p})e^{-{i(\vec{p}+\vec{k})\cdot\vec{x}}}
c^\dagger_r(\vec{k})d^\dagger_s(\vec{p})]~.&
\end{align*}
In the same way, we obtain $[\psi^\dagger_a(\vec{x})\psi_a(\vec{x}),\sum_{r}\int{d^3\vec{k}}d^\dagger_r(\vec{k})d_r(\vec{k})]$
\begin{align*}
=&\sum_{r}\int{d^3\vec{k}}\bigl(-\psi^\dagger_a(\vec{x})d^\dagger_r(\vec{k})
\{\psi_a(\vec{x}),d_r(\vec{k})\}+\{\psi^\dagger_a(\vec{x}),d^\dagger_r(\vec{k})\}d_r(\vec{k})
\psi_a(\vec{x})\bigr)&\\=&\frac{m^2}{(2\pi)^3}\sum_{s,r}\int
\frac{{d^3\vec{p}}{d^3\vec{k}}}{\sqrt{E_{\vec{p}}E_{\vec{k}}}}[
-u^\dagger_s(\vec{p})v_r(\vec{k})e^{-i(\vec{p}+\vec{k})\cdot\vec{x}}c^\dagger_s(\vec{p})
d^\dagger_r(\vec{k})&\\&-
v^\dagger_s(\vec{p})v_r(\vec{k})e^{i(\vec{p}-\vec{k})\cdot\vec{x}}d_s(\vec{p})d^\dagger_r(\vec{k})+
v^\dagger_r(\vec{k})u_s(\vec{p})e^{i(\vec{p}+\vec{k})\cdot\vec{x}}
d_r(\vec{k})c_s(\vec{p})&\\&+
v^\dagger_r(\vec{k})v_s(\vec{p})e^{-i(\vec{p}-\vec{k})\cdot\vec{x}}
d_r(\vec{k})d^\dagger_s(\vec{p})]~.&
\end{align*}
This can be simplified to
\begin{align*}
&[\psi^\dagger_a(\vec{x})\psi_a(\vec{x}),\sum_{r}\int{d^3\vec{k}}d^\dagger_r(\vec{k})d_r(\vec{k})]=&\\
-&\frac{m^2}{(2\pi)^3}\sum_{s,r}\int
\frac{{d^3\vec{p}}{d^3\vec{k}}}{\sqrt{E_{\vec{p}}E_{\vec{k}}}}[
u^\dagger_s(\vec{p})v_r(\vec{k})e^{-i(\vec{p}+\vec{k})\cdot\vec{x}}c^\dagger_s(\vec{p})
d^\dagger_r(\vec{k})]+
&\\
&\frac{m^2}{(2\pi)^3}\sum_{s,r}\int
\frac{{d^3\vec{p}}{d^3\vec{k}}}{\sqrt{E_{\vec{p}}E_{\vec{k}}}}[
v^\dagger_r(\vec{k})u_s(\vec{p})e^{i(\vec{p}+\vec{k})\cdot\vec{x}}
d_r(\vec{k})c_s(\vec{p})]~.&
\end{align*}
Putting the two results together, we get
\begin{align*}
&[\psi^\dagger_a(\vec{x})\psi_a(\vec{x}),N]=&\\
-&\frac{2m^2}{(2\pi)^3}\sum_{s,r}\int
\frac{{d^3\vec{p}}{d^3\vec{k}}}{\sqrt{E_{\vec{p}}E_{\vec{k}}}}[
u^\dagger_s(\vec{p})v_r(\vec{k})e^{-i(\vec{p}+\vec{k})\cdot\vec{x}}c^\dagger_s(\vec{p})
d^\dagger_r(\vec{k})]+
&\\
&\frac{2m^2}{(2\pi)^3}\sum_{s,r}\int
\frac{{d^3\vec{p}}{d^3\vec{k}}}{\sqrt{E_{\vec{p}}E_{\vec{k}}}}[
v^\dagger_r(\vec{k})u_s(\vec{p})e^{i(\vec{p}+\vec{k})\cdot\vec{x}}
d_r(\vec{k})c_s(\vec{p})]~,&
\end{align*}
which is not equal to zero, even if we think about fields as
distributions. If we start from the state
 $d^\dagger_s(p_0)c^\dagger_s(p_0)|0\rangle$, it is clear that there are well-behaved functions $f$ such that
\begin{equation*}
\langle 0|\int
d^3\vec{x}f(\vec{x})[\psi^\dagger(\vec{x})\psi(\vec{x}),N]|d^\dagger_s(p_0)c^\dagger_s(p_0)|0\rangle\neq
0~.
\end{equation*}
\vskip 30pt
\begin{eref}
\bibitem{bell1}
J. Bell, Beables for quantum field theory, {\em CERN-TH\/} {\bf
4035/84}, (1984).
\bibitem{bell2}
J. Bell, {\em Speakable and unspeakable in quantum mechanics \/},
(Cambridge university press, 1987).
\bibitem{bohm1}
D. Bohm, B. Hiley and P. Kaloyerou, An ontological basis for the
quantum theory. 2. A causal interpretation of quantum fields, {\em
Phys. Rept} {\bf 144} (1987) 323--375.
\bibitem{colin}
S. Colin, A deterministic Bell model, {\em Phys. Lett. A} {\bf 317} (2003) 349--358.
\bibitem{epr}
A. Einstein, B. Podolsky and N. Rosen, Can quantum-mechanical description of physical reality be considered complete?, 
 {\em Phys. Rev. \/} {\bf 47} (1935) 777--780.
\bibitem{hegerfeldt}
G. Hegerfeldt, Remark on causality and particle localization, {\em Phys. Rev.} {\bf D10} 
(1974) 3320--3321.
\bibitem{holland}
P. Holland, The quantum theory of motion, (Cambridge university
press, 2000).
\bibitem{teller}
P. Teller, An interpretive introduction to quantum field theory, (Princeton University Press, 1997).
\end{eref}
\end{document}